\documentclass[epj]{svjour}
\usepackage{graphicx}
\usepackage{graphics}
\usepackage{epsf}
\usepackage{amssymb}
\usepackage{amsmath}

\newcommand{\eq}{\begin{equation}}                                              
\newcommand{\eqe}{\end{equation}}                      

\begin{document}
\title{Two-phase flow model for energetic proton beam induced pressure 
waves in mercury target systems in the planned European Spallation 
Source}
\author{I. F. Barna$^1$, A. R. Imre$^1$, L. Rosta$^2$ and F. Mezei$^2$}
%
%
%
\institute
{$^1$Atomic Energy Research Institute (AEKI) of the Hungarian  
Academy of Sciences, \\ 
P.O. Box 49, H-1525 Budapest, Hungary, EU \\
$^2$ Research Institute for Solid State Physics and Optics of the Hungarian 
Academy of Sciences, \\  P.O. Box 49, H-1525 Budapest, Hungary, EU }

\date{Received: date / Revised version: date}


\abstract{Two-phase flow calculations are presented to investigate the thermo-hydraulical effects
of the interaction between 300 kJ proton pulses (2 ms long, 1.3 GeV) with a closed mercury loop which
can be considered as a model system of the target of the planned European Spallation
Source(ESS) facility. The two-fluid model consists of six first-order partial
differential equations that present one dimensional mass momentum and
energy balances for mercury vapour and liquid phases are capable
to describe quick transients like cavitation effects or shock waves. The absorption of the proton beam is
represented as instantaneous heat source in the energy balance equations.
Densities and internal energies of the mercury liquid-vapour system are calculated from
van der Waals equation, and the general method how to obtain such properties is also presented.
A second order accurate high-resolution shock-capturing numerical scheme
is applied with different kinds of limiters in the numerical calculations.
Our analysis show that even 75 degree temperature heat shocks cannot cause considerable cavitation effects 
in mercury in 2ms long pulses}

\maketitle


\section{Introduction}
\label{intro}
One non-destructive material research method is neutron scattering.
Free neutrons for neutron beams for research purposes need to be extracted from their
bound states of atomic nuclei. Energetic neutron beams can be produced in fission
of heavy elements (e.g. $^{235}U$) or by spallation.
In fission of $^{235}U$ 190 MeV heat is released for each extracted fast neutron 
while in the spallation process only about 30 MeV heat is deposited per fast neutron.
The deposited heat has to be removed by cooling and it ultimately becomes a limiting
thermodynamic factor for the amount of neutrons produced.
As a second distinct advantage of pulsed spallation sources over continuous ones is that
a larger part of the neutrons produced can be delivered to the sample in monochromatic
beams.
These two advantages of spallation sources make is possible to construct more powerful
neutron sources with larger neutron flux than ever before.
The simple goal of the planed European Spallation Source(ESS) is to provide Europe with the most
powerful neutron facility. A choice of a 5 MW proton beam power at 1.3 GeV proton energy
with 111 mA proton beam current and with 16.66 Hz repetition rate of 2 ms long neutron pulses will 
produce time average thermal neutron flux density of $3.1\times10^{14} n/cm^2 s$  in the ESS mercury target.
The proton pulse causes a thermal and a pressure shock in the target which may cause cavitation
or tensile stress. 
The question of cavitation erosion \cite{futak1} has crucial importance in the constructional planning
of any spallation neutron source target facility. A detailed analysis of the planned ESS can be
found elsewhere \cite{ess}.
In the following study we present and analyze a one-dimensional six-equation two-fluid
model which is capable to describe transients like pressure waves, quick vapour void fraction
creation and annihilation which is proportional to cavitation caused by energetic proton interaction
in mercury target.

Our model has a delicate numerical scheme and 
capable to capture shock waves and describe transient waves which may
propagate quicker than the local speed of sound \cite{izt}. Most of the two-phase models have numerical methods 
which describes usual flow velocities.  

Our model can successfully reproduce the experimental data of different one- or two-phase flow problems such
as ideal gas Riemann problem, critical flow of ideal gas in convergent-divergent nozzle, column separation or
cavitation induced water hammer, rapid depressurization of hot liquid from horizontal pipes or even steam condensation
induced water hammer \cite{waha}.

According to our knowledge there is no real two-phase flow calculation for mercury flow system.
Some timorous attempts were presented with the help of some commercial three dimensional industrial codes like
Fluent or ANSYS \cite{zurzi,05KOGISH} but the results are questionable. Some results show complete and immediate
vaporization during the first proton pulse, which is contradictory to experimental observations.
There are some studies for three dimensional numerical simulation of magnetohydrodynamic processes
in the muon colliders mercury target. These studies take strong external magnetic fields into
account \cite{samul} but consider single phase only, neglecting vaporization or condensations.
The liquid phase of mercury was modeled using the stiffened politropic equation of state and
the vapour phase was considered to be ideal gas.
There is a literature survey on various fluid flow data for mercury from the politropic equation state
\cite{cords} which can be directly applied in calculations.
There are also different equations of state (EOS) available
for mercury from microscopic molecular simulation \cite{kitamura,raabe} of from macroscopic
theories like virial expansion \cite{mehdipour} or from generalized van der Waals equation like the
Redlich-Kwong equation \cite{morita} or the like \cite{marti}.
Thermodynamical and flow properties of other liquid metals are also in the focus of
recent scientific interest \cite{nagr,morita}.

In the next sections we introduce our applied model, give a detailed analysis about phase transitions 
 and present pressure wave results.

\section{Theory}
\subsection{Theory of two-phase flow}

There is a large number of different two-phase flow models with different levels of
complexity \cite{stew,meni} which are all based on gas dynamics and shock-wave theory.
In the following we present our one dimensional six-equation equal-pressure two-fluid model.
The density, momentum and energy balance equations for both phases are the following:

\begin{eqnarray}     
\frac{\partial A (1-\alpha) \rho_l}{\partial t}  +
\frac{\partial A (1-\alpha) \rho_l (v_l-w) }{\partial x}  = -A \Gamma_g    \\
\frac{\partial A \alpha \rho_g}{\partial t} +
\frac{\partial A \alpha \rho_g (v_g-w) }{\partial x}  = A \Gamma_g  \\
\frac{\partial A(1- \alpha) \rho_l v_l}{\partial t} +  
\frac{\partial A(1- \alpha) \rho_l v_l(v_l-w) }{\partial x} + \nonumber \\ 
A(1-\alpha) \frac{\partial p}{\partial x} - A\cdot CVM   
- Ap_i\frac{\partial \alpha}{\partial x} =   AC_i|v_r|v_r \nonumber \\ 
- A\Gamma_g v_i + A(1-\alpha)\rho_l cos \theta - AF_{l,wall} \\ 
\frac{\partial A\alpha \rho_g v_g}{\partial t} +
\frac{\partial A \alpha \rho_g v_g(v_f-w) }{\partial x} +  
A\alpha \frac{\partial p}{\partial x} + \nonumber \\ A\cdot CVM + 
Ap_i\frac{\partial \alpha}{\partial x} = \nonumber \\ 
-AC_i|v_r|v_r  + A\Gamma_g v_i + A\alpha\rho_g cos \theta - AF_{g,wall}  
\end{eqnarray}  
\begin{eqnarray}
\frac{\partial A(1- \alpha) \rho_l e_l}{\partial t} +
\frac{\partial A(1- \alpha) \rho_l e_l(v_l-w) }{\partial x} + \nonumber \\
p \frac{\partial A(1- \alpha)}{\partial t} + \frac{\partial A(1- \alpha)
p (v_l-w) }{\partial x}  
= AQ_{il} \nonumber  \\ -A\Gamma_g(h_f+ v_l^2/2) + A(1-\alpha)\rho_l v_l g cos \theta  
+ E_{l,pulse}(x,t) \nonumber \\
\end{eqnarray}
\begin{eqnarray}
\frac{\partial A \alpha \rho_g e_g}{\partial t} +
\frac{\partial A \alpha \rho_g e_g(v_l-w) }{\partial x} + 
p \frac{\partial A \alpha}{\partial t} + \frac{\partial A \alpha   
p (v_l-w) }{\partial x} \nonumber \\
= AQ_{ig} +A\Gamma_g(h_g+ v_g^2/2) + A \alpha \rho_g v_g g cos \theta + E_{g,pulse}(x,t) 
\nonumber \\ 
\end{eqnarray}

Index l refers to the liquid phase and index g to the gas phase.
Nomenclature and variables are described in Table I.
Left hand side of the equations contain the terms with temporal and spatial derivatives.
Hyperbolicity of the equation system is ensured with the virtual mass term CVM and with the
interfacial term (terms with $p_i$).
Terms on the right hand side are terms describing the inter-phase heat, mass(terms with
$\Gamma_g$ vapour generation rate) volumetric heat fluxes $Q_{ig}$, momentum transfer
(terms with $C_i$), wall friction $F_{g_wall}$, and gravity terms.
A detailed analysis of the source terms can be found in \cite{waha}.
The last term in the energy equations $E_{i,pulse}(x,t)$ represents the deposited energy from the
proton beam and will be specified later on.

Two additional equations of state (EOS) are needed to close the system of equations (Eq. 1-6)
\eq
\rho_k = \left( \frac{\partial \rho_k}{\partial p} \right)_{u_k} dp +
 \left( \frac{\partial \rho_k}{\partial u_k} \right)_p du_k.
\eqe
Partial derivatives in Eq. 7 are expressed using pressure and specific internal energy
as an input.
In the following we show how the liquid-steam table - a sixfold numerical table -
($p,T,\rho_l, u_l, \rho_g, u_g$) can be created for mercury from an arbitrary EOS.
To avoid technical difficulties we do not modify (Eq. 1-6) including 
the used analytic EOS, just create a numerical liquid-steam table. 
In this manner arbitrary two phase-flow systems can be investigated with 
the same model in the future (e.g. lead-bismuth eutectic, liquid Li or He).
Liquid metal systems can operate on low (some bar) pressure and have larger
heat conductivity than water which can radically enhance thermal efficiency.
 
We start with the usually parameterized van der Waals EOS from
\eq
p = \frac{RT}{V-b} - \frac{a}{V^2}
\eqe
where R=8.314 J/mol/K is the universal gas constant and parameters a and b are related
with the critical molar volume ($V_c$) temperature ($T_c$) and pressure ($p_c$)
of the considered fluid:
$a = 9P_cV_c^2 \>\> b = V_c/3 $
For the critical temperature and pressure of Hg the $T_C = 1733 \pm 50 K$ and
$p_C = 160.8 \pm 5 MPa $ data were taken from \cite{morita}.
T, p and V are the temperature the pressure and the volume, respectively.
(We mention than in \cite{morita} the parameter $a (=9P_cV_c^2)$ is uncorectly given.) 
The fluid density with the corresponding saturated vapour density can be easily determined
from the EOS with the well-known Maxwell construction. To obtain the internal energies
for both phases is a bit more difficult task.
We start with the second law of thermodynamics
\eq
 du = Tds - pdV
\eqe
where s is the entropy and u is the internal energy.
With the Maxwell relations  $\left(\frac{\partial T}{\partial V}  \right)_s =
-\left(\frac{\partial P}{\partial s}  \right)_V  $   we end up with the following working equation
\eq
 du = c_VdT + \left[T\left(\frac{\partial p}{\partial T}  \right)_V -p\right] dV.
\eqe
The internal energy is a thermodynamical potential therefore the choice of the zero point
can be defined arbitrary, we took $T = 253.14 K $ which is 10 degree above the melting point of solid mercury
at normal pressure.
The heat capacity at constant volume $c_v$ may in turn be calculated from the heat capacity at constant
pressure $c_p $with the thermodynamic relation
\eq
c_p = c_V + T \left( \frac{\partial V}{\partial T}  \right)^2_p \left(\frac{\partial p}
{\partial V}  \right)_T
\eqe
where $ \frac{1}{V} \frac{\partial V}{\partial T} = \alpha_T$ is the thermal expansion coefficients.
(To avoid further misunderstanding in this study we use $\alpha_T$ for the thermal expansion
coefficient and $\alpha$ for vapour void fraction.)
Polynomial fits for the temperature dependence of experimental data  of heat capacity $c_p$
and expansion coefficient $\alpha_T$   \cite{lance}(or \cite{cords}) help us to calculate the internal
energy of the liquid state.
Finally, the internal energy of the corresponding gas phase has to be determined.
The critical temperate of mercury is at $T_C = 1733 \pm 50 K$. 
In the temperature range of 270-500 K (which is our recent interest)
the experimental heats of vapourization data \cite{raabe} can be satisfactory fitted
with a linear function.
With this method a two-phase liquid-steam table was constructed between 270-500 Kelvin
of temperatures and 7 to $7\cdot 10^7$ Pascal pressure.

Additional flow properties of mercury like dynamic viscosity and heat transfer coefficients are
approximated with piecewise continuous temperature dependent functions from \cite{cords}.
The surface tension was considered as a linear function of temperature \cite{jasp}.

The effect of the 300 kJ proton pulse was treated as a sudden thermal shock  which
means an additional source terms in both energy equations $E_{i,pulse}(x,t)$.
The deposited energy of the proton beam in the mercury target is proportional to the density.
With the introduction of the mixture density
$\rho_m = \alpha \rho_g + (1-\alpha)\rho_f $  the interaction between the proton-mercury
two-phase flow can be further improved.
According to experimental proton beam analysis the spatial energy distribution perpendicular to the
propagation direction has a parabolic shape \cite{ni,fut} with a diameter of 20 cm.
To describe well-defined finite duration we use $sin^2$ envelope with $\tau = 2ms$.
\eq
E_{g,pulse}(x,t) = \frac{\rho_g \alpha}{\rho_m}  E_0 sin^2\left(\frac{\pi t}{\tau} \right) (1- (x/x_s)^2)
\eqe
\eq
E_{f,pulse}(x,t) =  \frac{\rho_f (1-\alpha)}{\rho_m}  E_0 sin^2\left(\frac{\pi t}{\tau} \right) (1- (x/x_s)^2)
\eqe
The effective range of 1.3 GeV protons in mercury can be calculated with the 
Bragg theory and gives 41 cm \cite{srim}.
Experimental consideration state that 47 percent of the original 5 MW beam power is absorbed in the target
which is 2.37 MW. In the planned ESS facility a train of 16.66 proton pulses will come
with  2ms long pulse duration and the total sum of these pulses give the 5 MW beam power.
Hence, the peak power parameter $E_0$ has to be normalized in such a way that the spatial and time integral of
$E_{i,pulse}(x,t)$ gives the absorbed 2.37 MW power of the original 5 MW beam.
The system of Eqs. (1-6) represents the conservation laws and can be formulated
in the following vectorial form
\eq
{\bf{A}}\frac{\partial {\vec{\Psi}}} {\partial t} +  {\bf{B}}\frac{\partial {\vec{\Psi}}} {\partial x}  =
{\vec{S}} \label{equ}
\eqe
where ${\vec{\Psi}}$ represents the vector of the independent nonconservative variables
$ {\vec{\Psi}}(p,\alpha,v_f,v_g,u_f,u_g)$, ${\bf{A}}$ and ${\bf{B}}$ are the matrices of the system
and ${\vec{S}}$ is the source vector of non-differential terms. These three quantities ${\bf{A}}$, ${\bf{B}}$ and
${\vec{S}}$ can be obtained from Eq. (1-6) with some algebraic manipulation.
In this case the system eigenvalues which represent wave propagation velocities are given by the
determinant $det({\bf{B}}- \lambda{\bf{ A}})$.  An improved characteristic upwind discretization method is used to
solve the  hyperbolic equation system (\ref{equ}). The problem is solved with the combination of the first-
and second-order accurate discretization scheme by the so-called flux limiters to avoid numerical dissipation
and unwanted oscillations which appear in the vicinity of the non-smooth solutions. Exhaustive details about
the numerical scheme can be found in \cite{waha,izt}.

\subsection{Liquid-vapour phase transition in the metastable region}

Water boils at 100 Celsius (373.15 K) under atmospheric pressure; this is a well-known,
but not entirely correct piece of the common knowledge. Boiling is usually defined (at least
phenomenologically) when liquid-vapour phase transition happens not only at the already existing
interfaces, but within the bulk liquid too. For water, it happens usually at the already mentioned
100 Celsius, but not always. Overheating of liquids is a phenomenon, known for every chemistry students;
one can exceed the boiling point with a few degrees, without getting boiling, but then it can happen suddenly,
exploding the whole amount of liquid (and often the container too) \cite{deb}. In the following, we are going
 to explain this phenomenon and show its importance in the cavitation of mercury.
Liquid can co-exists with the vapour of the same material, without any problem.
The conditions (temperature and pressure) where they co-exist are described the vapour
pressure curve (also called saturation or co-existence curve). Liquid and vapour states can be
described by EOS; like van der Waals. A schematic isotherm (describing pressure
and volume on a constant temperature) can be seen on Figure 1/a. The isotherm has two extrema (marked as
B and D), these are the so-called spinodal points (liquid-vapour and vapour-liquid; LV and VL spinodals).
Between the two spinodals, the system would be unstable, due to the negative compressibility, therefore
these states (the ones on the curve between points B and D) cannot exist. The equilibrium conditions can
be calculated by using the Maxwell construction: a line (parallel to the V-axis) has to be drawn in a way
that the area between the isotherm and the Maxwell-line between points A and C and C and E has to be equal.
Then the intersects (A and E) gives the co-existing liquid and vapour volumes (or densities) and the
equilibrium pressure on the given temperature. Plotting the pressures on different temperatures, one would
obtain the vapour pressure curve, like the solid line on Figure 1/b. It can be seen that points A and E are
not special points of the isotherm. The liquid is not forced to boil at point A; it would be forced only at
point B (where liquid phase cannot exists any more). Plotting Bs at different temperature, one would obtain
the so-called liquid-vapour spinodal (dashed line on figure 1/b), the stability limit of liquid state.
The vapour-liquid spinodal (dot-dash line on figure 1/b) is important when we have over-saturated vapour;
we are going to neglect it here. The AB and DE parts on the isotherm are metastable; on Figure 1/b these
parts are represented by the region between the vapour pressure curve and the LV spinodal.
Liquid (without co-existing vapour phase) can exist in this region; even can exist under negative
pressure \cite{deb,tre}
The real boiling happens in this metastable region. Close to the vapour pressure curve the liquid is only
slightly metastable, can live for long time without nucleating vapour bubbles; far away from it
(close to the spinodal) the liquid will be very metastable and cavitate (boil) with higher probability.
The bubble nucleation can happen in two different ways. The heterogeneous nucleation happens when the
liquid already has a nucleus, usually a tiny bubble hidden in a crevice of the wall or stuck onto a floating
particle. Due to the small size (i.e. high curvature) the micro-bubble can be in equilibrium with a metastable
liquid for a while, but when the temperature is too high or the pressure is too low, it will initiate boiling.
The other process is the homogeneous nucleation. In that case, the initial micro-bubble will be produced by the
density fluctuations of the liquid; when the fluctuation is big enough to call it "bubble", then it will
initiate the boiling. In everyday life, boiling happens by heterogeneous nucleation, practically in the
immediate vicinity of the vapour pressure curve. In clean liquids (like distilled water) the boiling can
happen much farther. It is a well-known practice to avoid overheating (and the explosion-like boiling,
following it) to put some bubble seed into the liquid, like a few pieces of sponge-like pumice
(or boiling-) stones. In these nucleation processes -especially in homogeneous nucleation -
time is also an important factor; a liquid can endure high overheating/stretching for a small period of time
\cite{deb,tre,imrmar,imrmar2}. 
Therefore one cannot draw a well-defined line as nucleation
limit onto the phase diagram (Figure 1); the line drawn by us is only for demonstration.
the exact location depends on the purity of liquid, the amount of external disturbances
(even cosmic rays can generate bubbles in metastable liquids) and - in a great extent - on the time scale.
On Figure 1/c, a magnified part of Figure 1/b (without the VL spinodal, which is irrelevant in our case)
can be seen. K marks a state, where the liquid is in stable liquid phase; there is no vapour phase present.
To obtain phase transition, the temperature can be increased or the pressure has to be decreased.
By increasing the temperature (and keeping a constant pressure), the vapour pressure curve will be
reached at point L. This is the first point, where the liquid can boil and vapour phase might appear,
but in clear and undisturbed liquid, the probability of boiling here is very small. Increasing the
temperature further, the nucleation limit will be reached (point M); here the phase transition will surely
happen, due to heterogeneous or homogeneous nucleation, forming initially small, but continuously growing
separated bubbles, which later can merge into a continuous vapour phase. When the liquid is perfectly clean, all
disturbances are suppressed and the heating is very fast, etc., then this nucleation limit can be pushed
very close to spinodal limit (point N), where liquid phase cannot exist any more.
When the phase transition happens at the spinodal limit, one will obtain two
bi-continuous phases (liquid and vapour), instead of a continuous (liquid) and an
separated (vapour bubbles) one, obtained during nucleation. We should remark,
that for the first appearance of the vapour phase, the system will jump back to
the vapour pressure curve (which will be detected as a pressure jump).
Changing the pressure at constant temperature, one would reach the vapour pressure curve at point O, then
the nucleation limit at point P, finally the spinodal limit at point Q, with the same results as it
happens with temperature increase.
To see the extent of the effect, we will give numerical examples for water and for mercury. For water,
starting from room temperature (293.15 K, 1 bar), we will reach the vapour pressure curve
373.15 K. Increasing the temperature, boiling might happen any time; the highest experimentally
obtained value for overheating (i.e. point M) was around 570 K \cite{91ZHEDUR} giving almost 200 K overheat.
The spinodal temperature for water on atmospheric pressure is still debated, it has to be located
above the previously mentioned overheating limit, but certainly below the critical temperature of $1733 \pm 50$ K.
Also for water, by decreasing the pressure, the vapour pressure curve (point O) would be
reached at 0.025 bar pressure. The experimental limit of stretching is -1200 bar
\cite{91ZHEDUR}, where the estimated spinodal (depending on the model) is between -2000 and -4000 bar.
For mercury at 7 bar (which is the working pressure for the mercury in the ESS) the boiling point is at 760 K, it
is very far from the working temperature (which is close to room temperature, 300 K). The limit of overheating
is not known, but surely below the critical temperature, which is around 1700 K.
Concerning pressure drop, the vapour pressure of mercury around room temperature is almost zero (less than 2*10-6 bar);
concerning the fact that the working pressure is 7 bar, the possibility of a pressure drop of
this extent is very improbable. The measured nucleation limit of mercury at room temperature is in
the -2 to -425 bar range depending on purity\cite{53BRI}; therefore to get bubbles, the pressure should drop from +7 bar to
<-2 bar for a longer period. The absolute (spinodal) limit is unknown.
Although in the ESS, pressure decrease and temperature increase happens simultaneously, the working conditions
are so far from the beginning of the boiling region (vapour pressure curve)
that the possibility to reach it is negligible, except under special circumstances. First, there is a possibility
for fast pressure oscillation after the proton pulse; the amplitude 
can be even 300 bar \cite{01TALMOR},
which would be enough to cause cavitation. The other scenario would require gas-contamination (pre-existing gas
bubbles in the mercury); in that case even a tiny pressure decrease or temperature increase can cause the growing of
these micro-bubbles, mimicking boiling \cite{01TALMOR}. Non-uniform temperature and pressure distribution can cause
shear stresses, which can also cause cavitation in the liquid. Finally, the proton beam itself can initiate
cavitation, but only when the metastable states are already reached.
We can conclude, that with a few bar pressure drop and a few tens of K temperature increase, the cavitation
in the pure mercury target has low possibility. On the other hand, concerning
the reported cases of cavitation in similar facilities \cite{05DATFUT}
indicate, that either the conditions (T,p) might change more drastically or some phenomena,
neglected by us (like pressure oscillation, shear stresses, etc.) can play more important role. \\

\begin{figure}
 \vspace*{1.0cm}
 \hspace*{0.4cm}
\resizebox{0.8\columnwidth}{!}{
  \includegraphics{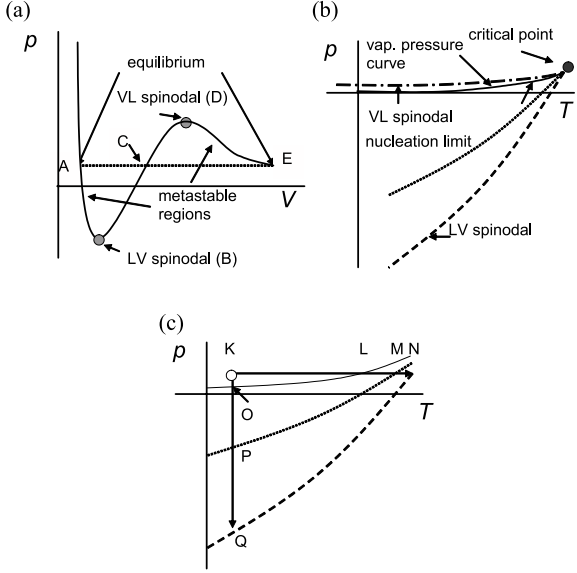}}
  \vspace*{2.4cm}
\caption{A subcritical isotherm of a van der Waals type fluid }
\label{fig:1}       
\end{figure}
\section{Results and discussion}   

The ESS mercury target loop is a complex facility with various pumps, heat exchangers and tanks \cite{ess}.
We model however with a simple six-sided closed loop (see Figure 1.) of a pipe with diameter of 5 cm and total length
of 5 m. The original temperature of the mercury is $T = 300$ K with pressure of 7 bar and flow velocity of $v= 4.6 $ m/s.
The proton beam interacts with a mercury via a $20 \times 5\> cm^2$   window. A simple calculation
shows that (47 \% of 300 kJ =) 141 kJ of energy will heat up 10 kg of mercury. The temperature jump 
of $\Delta t = 75 K$ is expected for ESS proton beam pulses. 
We applied a single pulse shot at time equal to zero and propagated Eqs. (\ref{equ}) to $t_{max} =4*10^{-2} $ sec.
A second order numerical scheme was used with the MINMOD flux limiter \cite{izt}.
For a satisfactory convergence the Courant number which measures the relative wave propagation speeds of the exact solution and the numerical
solutions was set to CFL = 0.6. The pressure history of the beam-target interaction point is presented in Fig 2.
After the initial pulse at $t = 1.6 $ms  the pressure reaches its maximal value which is 50 percent higher than the
original pressure. After the pulse the pressure does not fall below the initial pressure and
the temperature  will cool down to 300 K.
The mercury vapour void fraction was originally set to zero ($\alpha =10^{-12}$) which did not chang
during the time propagation allowing only "nanobubbles", too small to act as cavitation 
nuclei. If we consider one or two percent initial vapor void fraction 
(as a model for small bubbles) than a quick condensation can be observed. 
If we apply an elastic pipe with an elasticity of $2\times 10^{11} N/m^2$ Young's modulus 
(which are usual for steel) or/and include or exclude any kind of additional wall friction 
\cite{waha} for the fluid the pressure peaks will not be changed. This is clear fingerprint
 that the tube is still rigid enough.  
There is a strong indication that mercury is a non-wetting fluid on steel surface 
so the wall friction is negligible.  
	 		 
\begin{figure}
\resizebox{0.95\columnwidth}{!}{%
  \includegraphics{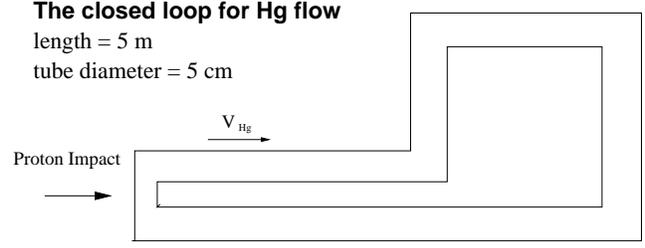}}
\caption{The schematic geometrical model of the ESS target }
\label{fig:1}       
\end{figure}


\section{Summary and outlook}
With the help of a one dimensional two-phase flow model we calculated the induced pressure
waves and vapour void fractions in mercury induced by energetic proton beams.
Our analysis showed that no vapour bubbles or cavitation effects can be seen after the
first absorbed proton pulse. Further, in depth analysis is in progress to investigate 
geometrical effects of the mercury target loop which is a complex facility with 
various pumps, heat exchangers and tanks \cite{ess}. 
Our model can include abrupt area changes, or convergent-divergent pipe cross section 
changes, or even heat exchangers.  
We modeled however with a simple six-sided closed loop (see Figure 1.) of a pipe with diameter of 5 cm and total length
of 5 m. The original temperature of the mercury is $T = 300$ K with pressure of 7 bar and flow velocity of $v= 4.6 $ m/s.
The proton beam interacts with a mercury via a $20 \times 5\> cm^2$   window. A simple calculation
shows that 141k J of energy will heat up 10 kg of mercury with a $\Delta t = 75 K$.
We applied a single pulse shot at time equal to zero and propagated Eqs. (\ref{equ}) to $t_{max} = 4*10^{-2} $ sec.
A second order numerical scheme was used with the MINMOD flux limiter \cite{izt}.
The Courant number which measures the relative wave propagation speeds of the exact solution and the numerical
solutions was set to CFL = 0.6. The pressure history of the beam-target interaction point is presented in Fig 3.
After the initial pulse at $t = 1.6 $ms  the pressure reaches its maximal value which is 50 percent higher than the
original pressure. After the pulse the pressure does not fall below the initial pressure and
the temperature  will cool down to 300 K.
The mercury vapour void fraction was originally set to zero ($\alpha =10^{-12}$) which 
did not changed during the time propagation allowing only "nanobubbles", too small to act as  
nucleus for cavitation. 
The question of the vapor void fraction, pipe elasticity or the liquid wall friction was examined 
also.

We would like to emphasize that further in-depth analysis is needed 
to clear up the question of a long pulse train.
The question of different equations of state will be investigated also.
As a long term interest we also planned to investigate other liquid metal
(e.g. bismuth-lead eutectic or liquid lithium) or liquid helium systems which can be 
interesting as a cooling media for new type of nuclear reactors. 
Liquid metal systems can operate on low (some bar) pressure
and have much larger heat conductivity than water which can radically 
enhance thermal efficiency.

\begin{figure}
\resizebox{0.95\columnwidth}{!}{%
  \includegraphics{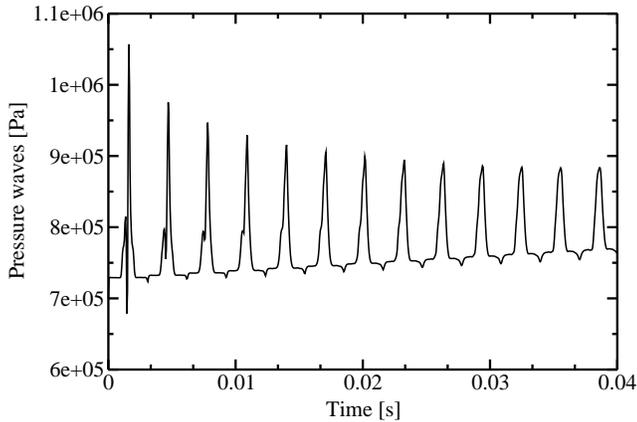}}
\caption{Time history of pressure at the point of the proton impact for the model in 
Fig. 2. with rigid tube walls.}
\label{fig:2}       
\end{figure}

   
\newpage 

\begin{table}
\caption{Nomenclature used in the two-phase flow equations(Eq. 1-6. )  }
\begin{center}
 \begin{tabular}{l} \hline \hline
  A   pipe cross section $(m^2) $ \\
      $C_i$ internal friction coefficient $(kg/m^4)$  \\
  CVM virtual mass term $(N/m^3)$ \\
   $e_i$ specific total energy [e = u + $v^2/2$] $(J/kg)$ \\
  $F_{f,wall}$ wall friction per unit volume $(N/m^3)$ \\
   g  gravitational acceleration$(m/s^2)$  \\
  $h_i$ specific enthalpy [h = u + $p/\rho$] $(J/kg)$ \\
   $p$ pressure (Pa) \\
   $p_i$ interfacial pressure $p_i = p\alpha(1-\alpha)$ (Pa) \\
  $Q_{ig}$ interf.-liq./gas heat transf. per vol. rate $(W/m^3)$ \\
     t time (s) \\
  $u_i$ specific internal energy $(J/kg)$  \\
     $v_i$ velocity $(m/s)$  \\
     $v_r$ relative velocity [$v_r = v_g -v_f$ ]$(m/s)$  \\
  w pipe velocity in flow direction $(m/s)$  \\
    x spatial coordinate (m) \\
  $\Gamma$  vapour generation rate $(kg/m^3)$ \\
    $\alpha$   vapour void fraction \\
   $\rho_i$ density $(kg/m^3)$ \\
    $\theta$  pipe inclination  \\ \hline \hline
  \end{tabular}
  \end{center}
\end{table}


\begin{thebibliography}{}
\bibitem{futak1} Futakava M., Naoe T., Tsai C.C., Kogawe H., Ishikura S., Ikeda Y., Soyama H., and
Date H. H. Cavitation Erosion in Mercury Target of Spallation Neutron Source
Fifth International Symposium on Cavitation (cav2003) Osaka, Japan, November 1-4, 2003

\bibitem{ess}  The European Spallation Source Project, Technical Report \\  $http://neutron.neutron-eu.net/n\_ess$




\bibitem{izt} Tiselj I. and Petelin S., 
J. Comp. Phys. {\bf{136}},  503-521 (1997).

\bibitem{waha}  Tiselj I., Horvath A., Cerne G., Gale J., Parzer I.,
Mavko B., Giot M., Seynhaeve J.M., Kucienska B. and Lemonnier H.   WAHA3 code manual,
Deliverable D10 of the WAHALoads project, March 2004

\bibitem{zurzi} 3rd High-Power Targetry Workshop, September 10-14, 2007 Bad Zurzach,
Switzerland $http://asq.web.psi.ch/hptrgts/index$

\bibitem{05KOGISH}   Hiroyuki Kogawa, {\it{et all.}}  
Journ. of Nucl. Mat. {\bf{34}}, 3178-183 (2005)


\bibitem{samul} Samuliak R.  Numerical simulation of hydro- and magneto-hydrodynamical properties
in the Muon Collider target. Lecture Notes in Comp. Sci, Vol. 2331 Springer-Verlag,
Berlin Heidelberg New York (2002) 391-400

\bibitem{cords} Cords H. A Literature Survey on Fluid Data for Mercury - Constitutive Equation  \\
 $http://neutron.neutron-eu.net/n\_ess$

\bibitem{kitamura} Kitamura H.,   
Journal of Chemical Physics {\bf{126}},  134509 (2007)

\bibitem{raabe} Raabe G. and Sadus R.J. 
Journ. of Chem. Phys. {\bf{119}},  6691 (2003)

\bibitem{mehdipour} Mehdipour N. and Bousheri A., 
Int. Journ. of Thermo-physics {\bf{18}},  1329 (1997)

\bibitem{morita} Morita K., Sobolev V. and Flad M., 
Journ.l of Nucl. Mat. {\bf{362}},  227-234 (2007)

\bibitem{marti} Martynyuk M.M.Z., Fiz. Khim. {\bf{65}}, 1716 (1981)

\bibitem{nagr} Mehdipour N., Boushehri A. and Eslami H., 
Journ. of Non-Cryst. Sol. {\bf{351}},  1333 (2005)

\bibitem{stew} Stewart H.B. and Wendroff B., 
J. Comp. Phys. {\bf{56}},  363 (1984)

\bibitem{meni} Menikoff R. and Plohr. B., 
Rev. Mod. Phys. {\bf{61}},  75-130 (1989)

\bibitem{lance} Davis L.A. and Gordon R.B.,  
Journ. Chem. Phys. {\bf{46}},   2650 (1967)

\bibitem{jasp} Jasper J.J.  
Phys. Chem. Ref. Data {\bf{1}},  841 (1972)

\bibitem{ni} Ni L., Bauer G.S. and Spitzer H., 
Nucl. Instr. Meth. in Phys. Res. A {\bf{425}},  57 (1999)

\bibitem{fut} Futakawa M., Kikuchi K., Conrad H. and Stechmesser H., 
Nucl. Instr. Meth. in Phys. Res. A {\bf{439}},  1 (2000)

\bibitem{srim} Ziegler J.F., Biersack J.P. and Littman U., The stopping 
and Ranges of Ions in Matter, Pergamon Press, Oxford (1985)

\bibitem{deb}   Debenedetti P.G. 1996 Metastable Liquids: Concepts and Principles 
 (Princeton University Press, Princeton)

\bibitem{tre} Trevena D.H. 1987 Cavitation and Tension in Liquids (Adam Hilger, Bristol)

\bibitem{imrmar} Imre A., Marti\'as K. and Rebelo L.P.N.
 J. Non-Equilib. Thermodyn.  {\bf{23}}, 351 (1998)

\bibitem{imrmar2}   Imre A R, Maris H J and Williams P R (Eds.) 
2002 Liquids Under Negative Pressure (NATO Science Series, Kluwer, Dordrecht)

\bibitem{53BRI}   Briggs L.J., 
J. Appl. Phys., {\bf{24}}, 488-490 (1953)

\bibitem{91ZHEDUR}     Zheng Q., Durben D.J., Wolf G.H. and Angell C.A. 
Science  {\bf{254}}, 829 (1991)

\bibitem{05DATFUT} Hidefumi D. and  Futakawa M., 
Int. Journ. Imp. Eng.  {\bf{32}}, 118-129 (2005)
 
\bibitem{01TALMOR} Taleyarkhan  R.P. and  Moraga F., 
Nucl. Eng. and Des. {\bf{207}}, 181-188 (2001)

\end{thebibliography}
\end{document}